\documentclass[12pt]{article}
\begin{document}
\baselineskip .3in
\begin{titlepage}
\begin{center}
{\large{\bf Composite Fermion Approach to Diquark and Heavy-Light
Baryon Masses }}\vskip .2in
  A.CHANDRA$^{\S}$, A. BHATTACHARYA $^{\ddag}$, B.CHAKRABARTI $^*$
\end{center}

\vskip .1in
\begin{center}
Department of Physics, Jadavpur University \\
Calcutta 700032, India.\\
\end{center}

\vskip .3in {\centerline{\bf Abstract}}
 A composite Fermion (CF) model of quasi particle has been used to
 describe a diquark. Considering baryons in quark-diquark
 configuration, the masses of the heavy light baryons like $\Lambda_{c}^{+}$, $\Sigma_{c}^{+}$,$\Xi_{c}^{0}$,
$\Omega_{c}^{0}$ and $\Lambda_{b}^{+}$,$\Sigma_{b}^{+}$,
$\Xi_{b}^{0}$, $\Omega_{b}^{-}$ have been computed using
 this CF model of the diquark. The results are found to be in
 good agreement with the corresponding experimental findings. It
has been suggested that the diquark can be well described in the
framework of CF model in a gauge invariant way.

 \vskip .3in
 $^{\ddag}$ e-mail: pampa@phys.jdvu.ac.in \vskip .2in
$^*$Permanent address: Department of Physics \\
Jogamaya Devi College, Kolkata, India.

e- mail:$ballari$\_$chakrabarti@yahoo.co.in$
\end{titlepage}

The role of diquark in baryon spectroscopy have been discussed by
number of authors $^{1-3}$. The diquark is supposed to be a
fundamental candidate for the structure and interaction of the
heavy baryons and exotics. The New LHC data opens up immense
possibility of identifying a numbers new particles which would
need to describe. Baryons described in the framework of
diquark-quark system describes the dynamics and interaction of the
baryons to a considerable extent particularly for heavy baryons.
Cakir et al$^{4}$ have investigated the
   resonant production of first generation scalar and vector diquarks in LHC collider and
   observed that LHC collider predicts a larger value of cross section. They have emphasized that the diquarks
   with different mass range could be investigated in LHC with suitable
   values of $\alpha$. Stability of such a correlated two quark system which has
been described as an independent object and which is supposed to
be an important component of the hadrons structure should be
studied with much attention. A number of models have been
suggested for diquarks $^{5,6}$.The possibility of forming
quark-quark and quark-antiquark system by Instanton Induced
Interaction(III) have been developed by Shuryak $^{7}$ and Schafer
et al $^{8}$. Betman et al $^{9}$ have investigated formation of
bound state of quark-quark or quark-antiquark systems due to
instanton induced interaction and predicted  that such bound state
is formed inside the hadron as a bubble of the size of the
instanton radius. The diquark has been described as a quasi
particle behaving like an independent entity by Bhattacharya et al
$ ^{10}$. Oka $^{11}$has made a detailed study on the diquark
correlation in the context of different models.

It has been suggested that the diquark as can be described in
gauge invariant way in the system of gauge interaction like two
dimensional electron gas in high magnetic field where electrons
can be described as composite Fermions(CF)$^{12}$. This in turn
may form Fermi liquid like state near the Fermi surface. Composite
fermion can have fractional charges and their spin is frozen. Such
CFs are described as the stable quasi particles in the system.
Raghavchari et al $^{13}$ have studied the quasi particle mass
which is fully gauge invariant and can be expressed as response
function of the system. In the present work we have used the CF
model for diquark describing it as a composite fermions as in the
work of Raghavchari et al $^{13}$ and have computed the masses of
diquarks. We have computed the masses of heavy-light baryon for
both charm and bottom sector. A good agreement with the
experimental results are obtained.

The effective mass of the composite can be defined in terms of the
self energy of composite fermions.In a fermi liquid, the low lying
excitations may be described as stable quasi particle and
quasi-hole excitation. These low energy eigen states can be
labelled as occupation configuration $n_{\overline{k}}$. Such a
state is smoothly connected to the corresponding state of the free
fermi system by adiabatically turning of interactions. The energy
of such states can be evaluated by Hellman-Feynman theorem
$^{14}$.The energy difference between ground state and the excited
state is related to the quasiparticle effective mass. Starting
from the Hamiltonian of a composite fermion with a momentum cut
off $\Lambda$ the expression for the quasi particle mass in a
gauge invariant system can be obtained as: $^{13}$ (with potential
V=0) :
\begin{equation}
\frac{1}{m^{*}}= \frac{1}{m}(1+ \frac{\Lambda^{4}}{2 k_{F}^{4}})
\end{equation}
 where $m^{*}$ is the effective mass of the CF , m is the mass of each
component, $k_{F}$ is the fermi momentum of the CF and $\Lambda$
is a cut off parameter.We have applied The CF picture for the
diquarks and the effective mass of diquark m$^{*}_{D}$ has been
expressed as ,
\begin{equation}
\frac{1}{m^{*}_{D}}=
\frac{1}{m_{q_{1}}+m_{q_{2}}}(1+\frac{\Lambda^{4}}{2 k_{F}^{4}})
\end{equation}
Where $m^{*}_{D}$ is the mass of the diquark, $m_{q_{1}}$,
m$_{q_{2}}$ are the constituent masses of the corresponding quark
flavours constituing the diquark. The fermi momentum of the
corresponding diquark has been estimated using the work of
Bhattacharya et al $^{15,16}$. In this work a relation between the
fermi momentum and the radius of a meson has been derived in the
frame work of the statistical model $^{15,16}$. We have used the
Fermi momentum of the meson consists of the same flavour of
diquark as the fermi momentum of the corresponding diquark. With
the input of the diquark radii from Castro et al ${17}$ we have
computed the Fermi momentum of diquarks and also the masses. The
masses of the diquarks using the expression (2) have been
furnished in Table-I.

Considering heavy baryons consists of a heavy quark, a diquark
consisting of light flavours and a suitable binding energy the
mass of heavy baryon can be expressed as:
\begin{equation}
M_{B}=m_{q}+m_{D}^{*}+E_{BE}
\end{equation}
where $m_{q}$ is the heavy quark mass, $m_{D}^{*}$ is diquark mass
and $E_{BE}$ is binding energy of the quark-diquark and has been
expressed as $E_{BE}$ = $<\psi |V| \psi>$. The potential has been
expressed as: V= b$r_{B}$ where b is the interaction parameter and
$r_{B}$ is the baryon radius. To estimate the binding energy we
have used the wave function from the statistical model$^{15,16}$.
In this model a baryon is assumed to be
 composed of a virtual $q\overline{q}$ in addition to the three valence
 quarks which determines the quantum number of the colourless
 baryon. The quarks real and virtual are assumed to be of same colour and flavour
 so that they may be regarded as identical and indistinguishable and are treated on the same
 footing. The indistinguishability of the valence quark with the
 virtual partner calls for the existence of quantum mechanical
 uncertainty in its available phase space. The valence quarks are
 assumed to be non- interacting with each other and
 considered to be moving almost independently in conformity with the
 experimental finding of asymptotic freedom. With above considerations we have come across
 an expression for the probability density of the baryons and the expression for $|\psi(r)|^{2}$ for a
baryon after normalization is obtained as,

\begin{equation}
|\psi(r)|^{2} = \frac{315}{64 \pi r_{B}^{9/2}}(r_{B}-r)^{3/2}
\Theta(r_{B}-r)
\end{equation}
where $r_{B}$ is the radius parameter of the baryon and $\Theta$
is the usual step function. We have estimated the masses of the
heavy light baryons using the expression (3).The results are
displayed in Table-II and Table-III for charm and bottom sector
respectively and are compared with the experimental findings
$^{18}$.

In the present work the masses of the heavy-light baryons
$\Lambda_{c}^{+}$, $\Sigma_{c}^{+}$,$\Xi_{c}^{0}$,
$\Omega_{c}^{0}$ and $\Lambda_{b}^{+}$,$\Sigma_{b}^{+}$,
$\Xi_{b}^{0}$, $\Omega_{b}^{-}$ have been computed using the
quark-diquark model of baryons considering the diquark as a
composite fermion. We have used cut off $\Lambda$ as 0.25 GeV
$^{19}$. The interaction parameter has been as b=0.3 GeV$^{2}$ as
in Lucha et al $^{20}$ for charm sector whereas for bottom sector
we have used b=1GeV$^{2}$ from the work of Liang et al $^{21}$.
The radii parameter of the baryons have been used from Brac et al
$^{22}$. It has been observed that for the charm sector very good
agreement with the experimental results have been obtained with
difference except $\Sigma_{c}^{+}$ where the difference is $\sim$
160 MeV. For the bottom sector the we have obtained little bit
lower value for $\Sigma_{b}^{+}$ and $\Omega_{b}^{-}$. It may be
pointed out that the some uncertainty may come from the radii of
the baryons and diquarks which are not exactly known and have a
substantial contribution in determining the fermi momentum and the
binding energy. It may be mentioned that this work gives us an
idea about the fermi momentum of the diqurks which is a very
important quantity for studying the properties of a particle.
 It has been suggested that the QCD vacuum can be
described as the choromo magnetic field. The CF which is bound
state of an electron and even number of vortices may have analogy
with the structure and formation of diquarks. In the present
investigation it has been observed the CF description of diquark
is  very successful in reproducing the masses of the baryons.
Although it is widely accepted that the diquark is building block
of baryons and exotics, the exact nature of it is yet to be known.
The diquark as CF may throw some light on the understanding of
structure and dynamics of the baryons which is quiet evident from
the present investigation. However we will investigate the problem
in our future work with non zero value of potential V.
 \vskip .3in
Acknowledgement; Authors are grateful to University Grants
Commission (U.G.C.), New Delhi, India for financial assistance
(Project Ref.No-F.No.37-217/2009(SR))

\newpage

\vskip 0.2in {\bf References}

\vskip .2in

\noindent [1].M. Anselmino et al., Rev. Mod. Phys. {\bf 65}, 1199
(1993).

\noindent [2]. E. V. Shuryak, Nucl. Phys. {\bf B 203}, 93 (1982).

\noindent [3]. A. S. Castro et al.,Z Phys. C {\bf 57}, 315 (1993).

\noindent [4]. O. Cakir et al.,arXiv:hep-ph/0508205.

\noindent [5]. M. Karlinar et al., Phys.Lett.{\bf 595}, 294
(2003).

\noindent [6]. R.L.Jaffe,Phys. Rev.Lett. {\bf 91},232003 (2003).

\noindent [7]. E. V. Shuryak, Nucl. Phys. {\bf B 50}, 93 (1982).

\noindent [8]. T.Schafer et al.,Rev. Mod. Phys.{\bf 70}, 323
(1998).

\noindent [9]. R.G.Betman et al.,Sov.J.Nucl.Phys.{\bf 41}, 295
(1985).

\noindent [10]. A. Bhattacharya et al.,Phys. Rev.C {\bf 81},
015202 (2010).

\noindent [11]. M.Oka,Prog.Theor.Phys. {\bf 112}, 1 (2004).

\noindent [12]. B.I.Helperin et al.,Phys.Rev.{\bf B 47}, 7312
(1993).

\noindent [13]. A. Raghavchari et al.,arXiv:cond.matt./9707055.

\noindent [14]D.J.Griffiths,"Introduction to Quantum
Mechanics",2nd Ed.,Pearson Edu.,(Singapore)Pte. Ltd,(2005),p-300.

\noindent [15]. A.Bhattacharya et al.,Int. J. Mod. phys. A {\bf
15}, 2053 (2003).

\noindent [16]. A. Bhattacharya et al.,Eur. Phys. J. C {\bf 2},671
(1998).

\noindent [17].A.S.De Castro et al., Z. Phys. C {\bf 57}, 315
(1993)

\noindent [18]. J.Beringer et al (PDG) Phys. Rev. D {\bf 86},
010001 (2012).

\noindent [19]. G.F.de Teramond, Hadrons and Strings, Trento,
July17-22, 2006.

\noindent [20]. W.Lucha et al.,Phys. Rep. {\bf 200}, 127 (1991).

\noindent [21]. H.Liang, arXiv:hep-ph/9807300.

\noindent [22]. B.S.Brac, Prog. Part. Nucl. Phys.,{\bf 36}, 263
(1996)

 \vskip .2 in

\newpage

\vskip 0.4in

TableI:Light-light Diquark Mass

 \vskip 0.4in
\begin{tabular}{|r |r |r r|}
\hline

 $Diquark$&$Momentum$&$Diquark-mass$&$( GeV)$\\
 $content$&$computed$&$Our$&$Other$\\
$[qq]_{0}$&$GeV$&$work $&$theory$\\

\hline
$[ud]_{0}$&$0.5880$&$0.708 $&$theory$\\
$[us]_{0}$&$0.6120$&$0.900 $&$theory$\\
$[ss]_{0}$&$0.6438$&$1.067 $&$theory$\\

\hline
\end{tabular}

TableII: Mass Spectrum (J$^{p}$=$\frac{1}{2}^{+}$) of the
Heavy-Light baryons (Charm Sector) .
 \vskip 0.4in
\begin{tabular}{|r r|r r|r r|r r|}
\hline

$\Lambda_{c}^{+}$&$  $&$\Sigma_{c}^{+}$& $  $&$\Xi_{c}^{0}$& $  $&$\Omega_{c}^{0}$&$  $\\

 $Theory$&$Exp$&$Theory$&$Exp$&$Theory$&$Exp$&$Theory$&$Exp$\\
 $(GeV)$&$(GeV)$&$(GeV)$&$(GeV)$&$(GeV)$&$(GeV)$&$(GeV)$&$(GeV)$\\

\hline
$  2.272$&$2.286  $&$2.292  $&$2.452 $&$2.464   $&$2.471   $&$2.636   $&$2.695\pm 0.0017    $\\

\hline
\end{tabular}

\vskip 0.3in
TableIII: Mass Spectrum (J$^{p}$=$\frac{1}{2}^{+}$)
of the Heavy-Light baryons (bottom Sector) .
 \vskip 0.4in
\begin{tabular}{|r r|r r|r r|r r|}
\hline
$\Lambda_{b}^{+}$&$  $&$\Sigma_{b}^{+}$& $  $&$\Xi_{b}^{0}$& $  $&$\Omega_{b}^{-}$&$  $\\

 $Theory$&$Exp$&$Theory$&$Exp$&$Theory$&$Exp$&$Theory$&$Exp$\\
 $(GeV)$&$(GeV)$&$(GeV)$&$(GeV)$&$(GeV)$&$(GeV)$&$(GeV)$&$(GeV)$\\

\hline
$ 5.496 $&$5.620  $&$5.551  $&$5.807 $&$5.707   $&$5.7924\pm0.003   $&$5.91   $&$6.165\pm0.0023    $\\

\hline
\end{tabular}
 \vskip 0.4in

\end{document}